\documentstyle[12pt]{article}
\newcommand{\beq}{\begin{equation}}
\newcommand{\eeq}{\end{equation}}
\newcommand{\bea}{\begin{eqnarray}}
\newcommand{\eea}{\end{eqnarray}}

\begin{document}
\setcounter{page}{0}
\topmargin 0pt
\oddsidemargin 5mm
\renewcommand{\thefootnote}{\fnsymbol{footnote}}
\newpage
\setcounter{page}{0}
\begin{titlepage}
\begin{flushright}
QMW 94-26
\end{flushright}
\begin{flushright}
hep-th/9408112
\end{flushright}
\vspace{0.5cm}
\begin{center}
{\large {\bf Conformal Invariance Of Interacting WZNW Models}} \\
\vspace{1.8cm}
\vspace{0.5cm}
{\large Oleg A. Soloviev
\footnote{e-mail: soloviev@V1.PH.QMW.ac.uk}\footnote{Work supported by the
PPARC and in part by a contract from the European Commission Human Capital
and Mobility Programme.}}\\
\vspace{0.5cm}
{\em Physics Department, Queen Mary and Westfield College, \\
Mile End Road, London E1 4NS, United Kingdom}\\
\vspace{0.5cm}
\renewcommand{\thefootnote}{\arabic{footnote}}
\setcounter{footnote}{0}
\begin{abstract}
{We consider two level $k$ WZNW models coupled to each other through a
generalized
Thirring-like current-current interaction. It is shown that in the large $k$
limit, this interacting system can be presented as a two-parameter
perturbation around a nonunitary WZNW model. The perturbation operators are the
sigma model kinetic terms with metric related to the
Thirring coupling constants. The renormalizability of the perturbed model leads
to an algebraic equation for couplings. This equation coincides with the master
Virasoro equation. We find that the beta functions of the two-parameter
perturbation have nontrivial zeros depending on the Thirring coupling
constants.
Thus we exhibit that solutions to the master equation
provide nontrivial conformal points to the system of two interacting WZNW
models.}
\end{abstract}
\vspace{0.5cm}
\centerline{August 1994}
 \end{center}
\end{titlepage}
\newpage
\section{Introduction}

At present, the importance of two dimensional conformal field theories (CFT's)
grows rapidly. It is being comprehended that the methods of CFT's
provide powerful tools to approach many physical problems from black
holes \cite{Witten-1},\cite{Callan} to turbulence
\cite{Polyakov-1},\cite{Polyakov-2}. Besides, CFT's continue to be a central
element of string theory.

Among all CFT's, those which admit a proper Lagrangian formulation appear to be
of an ultimate significance both in string theory and statistical physics. For
example, the minimal conformal models describing many statistical systems
\cite{Belavin},\cite{Friedan}, possess a very elegant Lagrangian
description in terms of
gauged WZNW models \cite{Karabali},\cite{Hwang}. Also, in string theory, many
exact string backgrounds are obtained from gauged WZNW models
\cite{Witten-1},\cite{Tseytlin-1}-\cite{Antoniadis}. These examples indicate
the
importance of Lagrangian CFT's formulated in terms of gauged WZNW models.

The interesting observation is that gauged WZNW models in turn can be
properly understood as ordinary WZNW theories
\cite{Novikov}-\cite{Knizhnik} coupled to each other through the (isoscalar)
Thirring-like
current-current interaction at a particular value of the Thirring coupling
constant \cite{Wiegmann}. However, the isoscalar coupling is the
simplest type of
general renormalizable current-current interactions. Hence, it is natural to
assume that at different values of Thirring coupling constants, the more
general Thirring current-current interaction ought to give rise to
new CFT's. It is
noteworthy that WZNW models with arbitrary current-current interaction emerge
naturally within the coadjoint orbit method \cite{Rai} as well as in the theory
of chiral WZNW models \cite{Bellucci}-\cite{Gates}. This makes the above
conjecture plausible.

The aim of the present paper is to shed some light on the issue of conformal
invariance of WZNW models with current-current interactions. We will consider
two unitary level $k$ WZNW models coupled to each other through a generic
Thirring-like current-current interaction. We will exhibit that there are
special values of the
Thirring coupling constants at which the interacting systems
can be understood as infrared conformal points of a certain nonconformal
theory. These critical values of the couplings will be shown to be connected to
solutions of the
master Virasoro equation \cite{Kiritsis},\cite{Morozov}. Thus we will give
more evidence for the existence of an intimate relation between Thirring models
and the affine-Virasoro construction \cite{Soloviev-1}.

The nonconformal theory whose infrared critical points we are going to study in
this paper is the
nonunitary WZNW model perturbed by two orthogonal quasimarginal relevant
operators. These perturbations are generalizations of the conformal operators
discussed in \cite{Soloviev-2}-\cite{Soloviev-4}. By establishing the link
between infrared conformal points of the perturbed nonunitary WZNW theory and
interacting unitary WZNW models, we will prove the conformal symmetry of the
latter.

The paper is organized as follows. In section 2 the definition of interacting
WZNW models is given. In section 3 the action of two interacting unitary level
$k$ WZNW models is presented in a factorized form which will be convenient for
implementing the $1/k$ method. In section 4 we will consider a two-parameter
perturbation around the nonunitary WZNW model. We will exhibit that at its
infrared conformal points the perturbed theory coincides with the system of two
interacting WZNW models with the coupling constants defined by solutions of the
master-Virasoro equation. Finally, in the last section we conclude with some
comments on the results obtained.

\section{Interacting WZNW models}

Let $S_{WZNW}(g_1,k)$ and $S_{WZNW}(g_2,k)$ be the actions of two conformal
level $k$ WZNW models. Let us consider the following interaction between these
two WZNW models\footnote{A
generalization to three and more interacting WZNW models is straightforward.}.
\begin{equation}
S(g_1,g_2,S)=S_{WZNW}(g_1,k)~+~S_{WZNW}(g_2,k)~+~S_I(g_1,g_2,S),\end{equation}
where these three terms respectively are given by
\begin{eqnarray}
S_{WZNW}(g_1,k)
&=&{-k\over4\pi}\left\{\int\mbox{Tr}|g^{-1}_1\mbox{d}g_1|^2~+~\frac{i}
{3}\int\mbox{d}^{-1}\mbox{Tr}(g_1^{-1}\mbox{d}g_1)^3\right\},\nonumber\\
S_{WZNW}(g_2,k)
&=&{-k\over4\pi}\left\{\int\mbox{Tr}|g^{-1}_2\mbox{d}g_2|^2~+~\frac{i}
{3}\int\mbox{d}^{-1}\mbox{Tr}(g_2^{-1}\mbox{d}g_2)^3\right\},\\
S_I&=&{-k^2\over\pi}\int d^2z~\mbox{Tr}^2(g_1^{-1}\partial g_1~S~\bar\partial
g_2g_2^{-1}),\nonumber\end{eqnarray}
with the coupling $S$ belonging to the direct product of two Lie algebras
${\cal G}_1\otimes{\cal G}_2$.
Here the fields $g_1$ and $g_2$ take their values in the Lie groups $G_1$ and
$G_2$ respectively. In what follows we will suppose $G_1=G_2=G$. The symbol
$\mbox{Tr}^2$ indicates a double tracing over the indices of a matrix from the
tensor product ${\cal G}_1\otimes{\cal G}_2$.

By construction, the classical theory in eq. (2.1) is invariant under conformal
transformations $z\to f(z),~\bar z\to\bar f(\bar z)$. Indeed, the coupling
matrix $S$ is dimensionless. Therefore, the classical interaction term is
conformally invariant by itself. In addition, there are affine symmetries
defined by the following transformations \cite{Soloviev-5}
\begin{equation}
g_1\to\bar\Omega_1(\bar z)g_1h_1\Omega_1(z)h_1^{-1},~~~~~~~~
g_2\to h_2^{-1}\bar\Omega_2(\bar z)h_2g_2\Omega_2(z),\end{equation}
where the functions $h_1,~h_2$ are computed from the equations
\begin{equation}
\bar\partial h_1h_1^{-1}=2k\mbox{Tr}~S~\bar\partial g_2g_2^{-1},~~~~~
h_2^{-1}\partial h_2=2k\mbox{Tr}~S~g_1^{-1}\partial
g_1.\end{equation}
Parameters $\Omega_{1,2}$ and $\bar\Omega_{1,2}$ are arbitrary functions of $z$
and $\bar z$ respectively.

The equations of motion of the theory described by the action in eq. (2.1)
coincide with the analiticity conditions of the local affine currents
\begin{equation}
\partial\bar{\cal J}_1=0,~~~~~~~~\bar\partial{\cal J}_2=0,\end{equation}
where
\begin{eqnarray}
\bar{\cal J}^{\bar a}_1&=&\bar J^{\bar a}_1~+~k\phi_1^{a\bar a}~S^{a\bar
b}~\bar
J^{\bar b}_2,\nonumber\\ & & \\
{\cal J}^a_2&=& J^ a_2~+~k\phi_2^{a\bar a}~S^{b\bar a}~
J^b_1.\nonumber\end{eqnarray}
Here we have used the following notations
\begin{eqnarray}
J^a_1&=&-\frac{k}{2}\mbox{Tr}(g_1^{-1}\partial g_1t^a),\nonumber\\
J^a_2&=&-\frac{k}{2}\mbox{Tr}(g_2^{-1}\partial g_2t^a),\nonumber\\
\bar J_1^{\bar a}&=&-\frac{k}{2}\mbox{Tr}(\bar\partial g_1g_1^{-1}t^{\bar
a}),\\
\bar J_2^{\bar a}&=&-\frac{k}{2}\mbox{Tr}(\bar\partial g_2g_2^{-1}t^{\bar
a}),\nonumber\\
\phi_1^{a\bar a}&=&\mbox{Tr}(g_1t^ag_1^{-1}t^{\bar a}),\nonumber\\
\phi_2^{a\bar a}&=&\mbox{Tr}(g_2^{-1}t^ag_2t^{\bar a}),\nonumber\end{eqnarray}
where $t^a$ are the generators of the Lie algebra ${\cal G}$ associated with
the Lie group $G$,
\begin{equation}
[t^a,t^b]=f^{abc}t^c,\end{equation}
with $f^{abc}$ the structure constants.

In general, the coupling matrix $S$ in eq. (2.1) can be chosen arbitrarily.
However, in what follows we will put restrictions on $S$ to be symmetrical and
invertible. In this case, the bosonic theory described by the action (2.1) is
equivalent to the non-Abelian fermionic Thirring model \cite{Soloviev-1}. Also,
it is interesting to point out that the equations
of motion (2.6) allow us to express any derivative of $J_{1,2},~\bar J_{1,2}$
in terms of the fields
$\phi_{1,2}$ and $J_{1,2},~\bar J_{1,2}$ without derivatives.

As we have mentioned above, the interaction between two WZNW models in eq.
(2.1) with arbitrary $S$ preserves the conformal invariance at the classical
level. However, in the course of quantization of the interacting theory, the
conformal symmetry can be broken. This can be seen on a simple example
$G=SU(2)$, $k=1$ and $S=\lambda~I$, where $I$ is unity in $su(2)\otimes su(2)$.
When $\lambda$ is small, the theory given by (2.1) is equivalent to the
sine-Gordon model \cite{Soloviev-6} which is not a conformal theory. At the
same time, there may exist some special values of the coupling matrix $S$ at
which the conformal invariance extends to the quantum level. The following
example illustrates such a possibility. Let us take $S=I/2k$. With the given
$S$ one can use the Polyakov-Wiegmann formula \cite{Wiegmann} to obtain
\begin{equation}
S(g_1,g_2,S=\frac{I}{2k})=S_{WZNW}(f,k),\end{equation}
where $f=g_1g_2$. The conformal invariance of this theory follows immediately
from the fact that the right hand side of eq. (2.9) is a conformal WZNW model.
Of course, the Polyakov-Wiegmann conformal point is very peculiar because it
has the extra gauge invariance under $g_1\to
g_1\Lambda,~g_2\to\Lambda^{-1}g_2$, where $\Lambda$ is arbitrary nondegenerate
matrix. Nevertheless, the considered example signifies that nontrivial
conformal points in the theory (2.1) may actually exist. The rest of the paper
will be dedicated to clarifying this conjecture.

\section{Expansion in the couplings}

Let us assume that the coupling matrix $S$ can be presented in the following
form
\begin{equation}
S=\sigma~\hat S,\end{equation}
where $\sigma$ is a small parameter. Apparently, the interaction term in eq.
(2.1) is linear in $\sigma$. We will show that this theory can be, in fact,
recast in a new form without interaction between $g_1$ and $g_2$ but instead
with a highly nonlinear dependence on $\sigma$.

Let us make in the theory (2.1) the following change of variables
\begin{eqnarray}
g_1&\to&\tilde g_1,\nonumber\\ & & \\
g_2&\to&h(\tilde g_1)\cdot \tilde g_2,\nonumber\end{eqnarray}
where the function $h(\tilde g_1)$ is the solution of the following equation
\begin{equation}
\partial h\cdot h^{-1}=-2k\sigma~\mbox{Tr}~\hat S\tilde g_1^{-1}\partial \tilde
g_1.\end{equation}
Since $h$ is a nonlocal function of $\tilde g_1$, the functional of the new
variable $\tilde g_1$ is going to be nonlocal as well.

One can check that the action given by eq. (2.1) in the new variables $\tilde
g_1,~ \tilde g_2$ takes the form
\begin{eqnarray}
S(g_1,g_2,k)&\to& S(\tilde g_1,\tilde g_2,k)
=S_{WZNW}(\tilde g_2,k)~+~S_{WZNW}(\tilde g_1,k)\nonumber\\ & &\\
&+&S_{WZNW}(h,k)
{}~-~\frac{k^2\sigma}{\pi}~\int
d^2z~\mbox{Tr}^2~\tilde g_1^{-1}\partial \tilde g_1~\hat S~\bar\partial
hh^{-1}.\nonumber\end{eqnarray}
Thus, after the change, the new field $\tilde g_2$ completely decouples from
$\tilde g_1$.\footnote{The important point to be made is that the Jacobian of
the change of variables in eq. (3.11) is equal to one. Therefore, the
factorized action given by eq. (3.13) holds at the quantum level as well.}
The price we pay for this factorization is a highly
nonlocal theory of the variable $\tilde g_1$. The last theory is described by
the following nonlocal functional
\begin{equation}
S(\tilde g_1)=S_{WZNW}(\tilde g_1,k)~+~S_{WZNW}(h(\tilde g_1),k)
{}~-~\frac{k^2\sigma}{\pi}~\int
d^2z~\mbox{Tr}^2~\tilde g_1^{-1}\partial \tilde g_1~\hat S~\bar\partial
h(\tilde
g_1)h^{-1}(\tilde g_1).\end{equation}
The given model is very complicated, because of the function $h$. The latter
can
be found by iterations from eq. (3.12). Moreover, the final result can be
expressed
in terms of the functions $\tilde J_1,~\bar{\tilde J_1}$ and $\tilde\phi_1$.
There is a useful formula
\begin{eqnarray}
&&\bar\partial hh^{-1}(z,\bar z)=-2k\sigma~\int d^2y~\bar\partial_{\bar
z}G(z,y)~\{\mbox{Tr}~\hat S\tilde g_1^{-1}(y,\bar
y)\partial_y\left(\bar\partial_{\bar y}\tilde g_1(y,\bar y)
\tilde g_1^{-1}(y,\bar
y)\right)\tilde g_1(y,\bar y)\nonumber\\ & & \\
&+&[\mbox{Tr}~\hat S\tilde g_1^{-1}(y,\bar
y)\partial_y\tilde g_1(y,\bar y),\bar\partial h(y,\bar y)h^{-1}(y,\bar
y)]\}.\nonumber\end{eqnarray}
{}From this equation, one can express $\bar\partial hh^{-1}$ in terms of
$\tilde
g_1$. The Green function $G(z,y)$ in eq. (3.15) satisfies the
following relation
\begin{equation}
\bar\partial\partial G(z,y)=\delta^{(2)}(z,y).\end{equation}

After substitution of the expression given by eq. (3.15) into the functional
(3.14), we find
\begin{equation}
S(\tilde g_1)=S_{WZNW}(\tilde g_1,k)~-~\frac{k^3\sigma^2}{\pi}~\int
d^2z~\mbox{Tr}\left(\mbox{Tr}~\hat S\tilde g_1^{-1}\partial \tilde
g_1\cdot\mbox{Tr}~\hat S\tilde g_1^{-1}\bar\partial \tilde g_1\right)~+~{\cal
O}(\sigma^3).\end{equation}
In formula (3.17), we have used the equations of motion (2.5) in order to
obtain the expression only in terms of the fields
$\tilde\phi_1$ and $\tilde J_1,~\bar{\tilde J}_1$. Indeed, eq. (3.17) can be
rewritten as follows
\begin{equation}
S(\tilde g_1)=S_{WZNW}(\tilde g_1,k)~-~\tau~\int d^2z~\Sigma_{ab}\tilde
J_1^a\bar{\tilde J_1^{\bar b}}\tilde\phi_1^{b\bar b}~+~{\cal
O}(\sigma^3),\end{equation}
where we have introduced the following notations
\begin{equation}
\tau=\frac{16k\sigma^2}{\pi},~~~~~~~~~
\Sigma_{ab}=\hat S^{a\bar a}\hat S^{b\bar
a}.\end{equation}

The obtained representation for $S(\tilde g_1)$ in eq. (3.18) looks like an
expansion in the coupling $\sigma$
around the conformal theory described by the level $k$ WZNW model
$S(\tilde g_1,k)$. However, we cannot go on with the given perturbation of the
unitary WZNW model, because the
operator $L_{ab}\tilde J_1^a\bar{\tilde J_1^{\bar b}}$ is irrelevant
around this conformal model. In the next section, we will show
that the perturbed theory in eq. (3.18) at special values of the Thirring
coupling matrix $S_{a\bar a}$ can be understood as the infrared limit of a
two-parameter perturbation around another conformal model in which the
mentioned above operator becomes relevant.

\section{Two-parametrical perturbation around the nonunitary WZNW model}

Given a conformal level $l$ WZNW model, we can build up the following composite
operator \cite{Soloviev-4}
\begin{equation}
O^{L,\bar L}=L_{ab}\bar L_{\bar a\bar b}~:J^a\bar J^{\bar a}\phi^{b\bar
b}:,\end{equation}
where
\begin{eqnarray}
J&=&J^at^a=-\frac{l}{2}g^{-1}\partial g,\nonumber\\
\bar J&=&\bar J^at^a=-\frac{l}{2}\bar\partial gg^{-1},\\
\phi^{a\bar a}&=&\mbox{Tr}:g^{-1}t^agt^{\bar a}:.\nonumber\end{eqnarray}
The product of the three operators in eq. (4.20) is defined according to
\cite{Soloviev-2}-\cite{Soloviev-4}
\begin{equation}
O^{L,\bar L}(z,\bar z)=L_{ab}\bar L_{\bar a\bar b}~\oint\frac{dw}{2\pi i}\oint
\frac{d\bar w}{2\pi i}{J^a(w)\bar J^{\bar a}(\bar w)\phi^{b\bar b}(z,\bar
z)\over|z-w|^2},\end{equation}
where the product in the numerator of the integrand is understood as an OPE. It
is easy to see that the given product does not contain singular terms provided
the matrices $L_{ab}$ and $\bar L_{\bar a\bar b}$ are symmetrical
\footnote{Indeed, the field $\phi$ is an affine primary vector. Therefore, its
OPE with the affine current $J$ is as follows
\begin{eqnarray}
J^a(w)\phi^{b\bar b}(z,\bar z)={f^{abc}\over(w-z)}\phi^{c\bar b}(z,\bar
z)~+~regular~terms.\nonumber\end{eqnarray}
Substituting this formula in eq. (4.22), one can see that only regular terms
will contribute provided $L_{ab}$ is a symmetrical matrix.}.

By definition, the operator $O^{L,\bar L}$ is an affine descendant of $\phi$.
Indeed, $O^{L,\bar L}$ can be presented in the form
\begin{equation}
O^{L,\bar L}(0)=L_{ab}\bar L_{\bar a\bar b}~J^a_{-1}\bar J^{\bar
a}_{-1}\phi^{b\bar b}(0),\end{equation}
where
\begin{equation}
J_m^a=\oint\frac{dw}{2\pi i}~w^mJ^a(w),~~~~~~\bar J_m^a=\oint\frac{d\bar w}
{2\pi i}~\bar w^m\bar J^a(\bar w).\end{equation}
Being an affine descendant, the operator $O^{L,\bar L}$ continues to be a
Virasoro primary operator. Indeed, one can check that the state $O^{L,\bar
L}(0)|0\rangle$ is a highest weight vector of the Virasoro algebra, with
$|0\rangle$ the $SL(2,C)$ invariant vacuum. That is
\begin{equation}
L_0~O^{L,\bar L}(0)|0\rangle=\Delta_O~O^{L,\bar L}(0)|0\rangle,~~~~~~
L_{m>0}~O^{L,\bar L}(0)|0\rangle =0.\end{equation}
Here the generators $L_n$ are given by the contour integrals
\begin{equation}
L_n=\oint \frac{dw}{2\pi i}~w^{n+1}T(w),\end{equation}
where $T(w)$ is holomorphic component of the affine-Sugawara stress-energy
tensor of the conformal WZNW model,
\begin{equation}
T(z)={:J^a(z)J^a(z):\over l+c_V}.\end{equation}
In eqs. (4.25), $\Delta_O$ is the conformal dimension of the operator
$O^{L,\bar L}$. It is not difficult to find that
\begin{equation}
\Delta_O=\bar\Delta_O=1~+~{c_V\over l+c_V}.\end{equation}
Here $\bar\Delta_O$ is the conformal dimension of $O^{L,\bar L}$ associated
with antiholomorphic conformal transformations. The quantity $c_V$ is defined
according to
\begin{equation}
f^{acd}f^{bcd}=c_V~\delta^{ab}.\end{equation}

{}From the formula for anomalous conformal dimensions of the operator
$O^{L,\bar
L}$, it is clear that when $l$ is negative large (namely, when $-l>c_V$), the
conformal dimensions are
in the range between 0 and 1, $0<\Delta_O=\bar\Delta_O<1$. Hence, in this
limit,
the operator $O^{L,\bar L}$ becomes a relevant conformal operator.
Correspondingly, for positive $l$ the operator $O^{L,\bar L}$ is irrelevant.

In spite of
the nonunitarity of the conformal WZNW model with negative level, the operator
$O^{L,\bar L}$
corresponds to a unitary highest weight vector of the Virasoro algebra.
Indeed, as we have shown above, $O^{L,\bar L}$ has positive conformal
dimensions, whereas the Virasoro central charge of the nonunitary WZNW model in
the large negative level limit is greater than one,
\begin{equation}
c_{WZNW}(l)={l\dim G\over l+c_V}=\dim G~+~{\cal O}(1/l)>1.\end{equation}
Thus, the operator $O^{L,\bar L}$ lies in the unitary range of the Kac
determinant and, hence, it provides a unitary representation of the Virasoro
algebra.

Clearly, operators $O^{L,\bar L}$ with arbitrary symmetrical matrices
$L_{ab},~\bar L_{\bar a\bar b}$ are Virasoro primary vectors with the same
conformal dimensions. However, their fusion algebras may be different. Among
all given operators, there are operators which obey the following fusion
algebra \cite{Soloviev-4}
\begin{equation}
O^{L,\bar L}\cdot
O^{L,\bar L}=\left[O^{L,\bar L}\right]~+~\left[I\right]~+~...,\end{equation}
where the square brackets denote the contributions of $O^{L,\bar L}$ and
identity operator $I$ and their descendants, whereas dots stand for all other
admitted operators with different conformal dimensions.

We have discovered \cite{Soloviev-4} that the fusion given by eq. (4.32)
results in the so-called master Virasoro equations
\cite{Kiritsis},\cite{Morozov} for the matrices  $L,~\bar L$. For example, for
the matrix $L$ this equation is formulated as follows
\begin{equation}
L_{ab}=L_{ac}L_{cb}~+~\frac{2}{l}\left(L_{cd}L_{ef}f^{cea}f^{dfb}~+~L_{cd}
f^{cdf}f^{dfa}L_{be}~+~L_{cd}f^{cef}f^{dfb}L_{ae}\right).\end{equation}
The matrix $\bar L$ satisfies a similar equation. Note that the master equation
has been originally obtained in the course of investigation of embeddings of
the
affine-Virasoro constructions into the affine algebra
\cite{Kiritsis},\cite{Morozov}. Now we found that the same equation emerges in
the fusion algebra of the {\it nonunitary} WZNW model. Due to the
interesting flip of sign of the level in the fusion algebra, the unitary WZNW
model gives rise to the master equation corresponding to a nonunitary
affine-Virasoro construction \cite{Soloviev-4}.

In what follows, we will deal with the large $|l|$ limit. Therefore, it is
interesting to look at the operator $O^{L,\bar L}$ in this limit. In the WZNW
model, the given limit is the classical limit. This allows us to use
classical expressions for the operators $J,~\bar J$ and $\phi$. We find
\begin{equation}
O^{L,\bar L}\to G_{\mu\nu}~\partial x^\mu\bar\partial
x^\nu~+~B_{\mu\nu}~\partial x^\mu\bar\partial x^\nu,\end{equation}
where $x^\mu$ are coordinates on the group manifold $G$, whereas
\begin{equation}
G_{\mu\nu}=-\frac{k^2}{8}L_{ab}\bar L_{\bar a\bar b}~\phi^{b\bar
b}e^a_{(\mu}\bar e^{\bar a}_{\nu)},~~~~~~
B_{\mu\nu}=-\frac{k^2}{8}L_{ab}\bar L_{\bar a\bar b}~\phi^{b\bar
b}e^a_{[\mu}\bar e^{\bar a}_{\nu]}.\end{equation}
Here $e^a_\mu$ and $\bar e^{\bar a}_\mu$ define left- and right-invariant
Killing vectors respectively. Note that when $L=\bar L$, the antisymmetric
field $B_{\mu\nu}=0$.

Thus, in the classical limit ($|l|\to\infty$), the operator $O^{L,\bar L}$
becomes the nonlinear sigma model. Therefore, the renormalizability of the
sigma model together with the master equation will guarantee the absence of
dots on the right hand side of eq. (4.32) in the large $|l|$ limit.

Now we are going to demonstrate that given a solution of the master equation,
we can
construct four orthogonal to each other operators of the $O^{L,\bar L}$ type.
It is known that if $L_{ab}$ is a solution of the master equation, then the
matrix
\begin{equation}
\tilde L_{ab}={l\delta^{ab}\over l-2c_V}~-~L_{ab}\end{equation}
is also a solution of the same master equation \cite{Kiritsis}. Similarly, this
will be true for $\bar L$. Let us show that the operator $O^{L,\bar L}$ is
orthogonal to the operator $O^{\tilde L,\bar L}$, where $\tilde L$ is given by
eq. (4.35). By computing the two point function, we find
\begin{equation}
\langle O^{L,\bar L}(1)O^{\tilde L,\bar L}(0)\rangle = {l^2\over4\dim
G}\left(L_{aa}~-~L_{ab}L_{ab}~+~\frac{2}{l}L_{ab}L_{cd}f^{ack}f^{bdk}\right)
\bar L_{\bar a\bar a}.\end{equation}
But the expression in parenthesis vanishes due to the master equation. Thus, we
arrive at
\begin{equation}
\langle O^{L,\bar L}(1)O^{\tilde L,\bar L}(0)\rangle =0.\end{equation}
In the same fashion, one can prove that there are two more operators, namely,
$O^{L,\bar{\tilde L}}$ and $O^{\tilde L,\bar{\tilde L}}$, which are orthogonal
both to each other and to the two operators just considered above.
Obviously, all the four operators share the same dimensions and the similar
fusion algebras. Therefore, all of them are appropriate for performing
renormalizable perturbations around the nonunitary WZNW model. However, for our
purposes, it will be sufficient to use only two operators $O^{L,\bar L}$ and
$O^{\tilde L,\bar L}$. It is necessary to point out that these operators are
conjugated to each other under the following transformation
\begin{equation}
L\to{lI\over l-2c_V}~-~L.\end{equation}

Now we define a new theory
\begin{equation}
S(\epsilon,\tilde\epsilon)=S_{WZNW}(g,l)~-~\epsilon~\int d^2z~O(z,\bar
z)~-~\tilde\epsilon~\int d^2z~\tilde O(z,\bar z),\end{equation}
where $\epsilon$ and $\tilde\epsilon$ are small parameters. From now on, we
will
omit superscripts on the perturbation operators.

We proceed to calculate the renormalization beta functions associated with the
couplings $\epsilon,~\tilde\epsilon$. Away of criticality, where
$\epsilon\ne0,~\tilde\epsilon\ne0$, the renormalization group equations are
given by \cite{Cardy}
\begin{eqnarray}
\frac{d\epsilon}{dt}\equiv\beta&=&(2-2\Delta_O)\epsilon~-~\pi
C\epsilon^2~+~{\cal
O}^3(\epsilon,\tilde\epsilon),\nonumber\\ & & \\
\frac{d\tilde\epsilon}{dt}\equiv\tilde\beta&=&(2-2\Delta_O)\tilde\epsilon~-~\pi
\tilde C\tilde\epsilon^2~+~\tilde{\cal
O}^3(\epsilon,\tilde\epsilon),\nonumber\end{eqnarray}
where the coefficients $C,~\tilde C$ are to be computed from the equations
\begin{eqnarray}
\langle O(z_1,\bar z_1)O(z_2,\bar z_2)O(z_3,\bar
z_3)\rangle&=&C~||O||^2~\Pi^3_{i<j}{1\over|z_{ij}|^{2\Delta_O}},\nonumber\\ & &
\\
\langle \tilde O(z_1,\bar z_1)\tilde O(z_2,\bar z_2)\tilde O(z_3,\bar
z_3)\rangle&=&\tilde C~||\tilde O||^2~\Pi^3_{i<j}{1\over|z_{ij}|^{2\Delta_O}},
\nonumber\end{eqnarray}
with $z_{ij}=z_i-z_j$. Here
\begin{equation}
||O||^2=\langle O(1)O(0)\rangle,~~~~~~~||\tilde O||^2=\langle \tilde O(1)\tilde
O(0)\rangle.\end{equation}

By using results of \cite{Soloviev-3},\cite{Soloviev-4}, one can compute the
coefficients $C,~\tilde C$. We will be interested only in leading orders in
$1/l$. To leading orders in $1/l$, the coefficients are given by
\begin{equation}
C={F^{abc}f^{abc}\bar F^{\bar a\bar b\bar c}f^{\bar a\bar b\bar c}\over
c_VL^{(0)}_{dd}\bar L^{(0)}_{\bar d\bar d}}~+~{\cal O}(1/l),~~~~~~~
\tilde C={\tilde
F^{abc}f^{abc}\bar F^{\bar a\bar b\bar c}f^{\bar a\bar b\bar c}\over
c_V\tilde
L^{(0)}_{dd}\bar L^{(0)}_{\bar d\bar d}}~+~\tilde {\cal
O}(1/l).\end{equation}
Here
\begin{eqnarray}
F^{abc}&=&f^{lmn}L^{(0)}_{al}L^{(0)}_{lm}L^{(0)}_{cn},\nonumber\\
\bar F^{\bar a\bar b\bar c}&=&f^{\bar l\bar m\bar n}\bar L^{(0)}_{\bar a\bar l}
\bar L^{(0)}_{\bar l\bar m}\bar L^{(0)}_{\bar c\bar n},\\
\tilde F^{abc}&=&f^{lmn}(\delta_{al}-L^{(0)}_{al})(\delta_{bm}-L^{(0)}_{
bm})(\delta_{cn}-L^{(0)}_{cn}),\nonumber\end{eqnarray}
where $L^{(0)},~\bar L^{(0)}$ are coefficients in the expansions of
$L,~\bar L$ in $1/l$:
\begin{equation}
L_{ab}=L^{(0)}_{ab}~+~{\cal O}(1/l),~~~~~~~
\bar L_{\bar a\bar b}=\bar L^{(0)}_{\bar a\bar b}~+~\bar{\cal O}(1/l^2).
\end{equation}
The expressions for $L^{(0)},~\bar L^{(0)}$ can be obtained from
the master equation. It has been found \cite{Obers}
\begin{equation}
L^{(0)}_{ab}=\sum_c~\Omega^{(0)}_{ac}\Omega^{(0)}_{bc}\theta^c.\end{equation}
There is a similar representation for $\bar L^{(0)}$. The quantities
$\Omega^{(0)},~\theta$ are defined as follows
\begin{equation}
\Omega^{(0)}{\Omega^{(0)}}^T=1,~~~~~~~~
\theta^a=0~~\mbox{or}~~1,~~~~~~~a=1,...,\dim G.
\end{equation}
Besides, the quantities $\theta^a,~\Omega^{(0)}_{bc}$ are subject to the
following quantization conditions \cite{Obers}
\begin{equation}
0=\sum_{cd}~\theta^c(\theta^a+\theta^b-\theta^d)\hat f^{cda}\hat
f^{cdb},~~~~~a<b,\end{equation}
where
\begin{equation}
\hat f^{abc}=f^{lmn}\Omega^{(0)}_{al}\Omega^{(0)}_{bm}\Omega^{(0)}_{cn}.
\end{equation}

Thus, given a solution of the master equation, we can derive
the constants $C$ and $\tilde C$ and, correspondingly, fixed points
$\epsilon^{*},~\tilde\epsilon^{*}$ of the beta functions in eqs. (4.40). It is
not difficult to find
\begin{equation}
\epsilon^{*}=-{2c_V\over\pi Cl},~~~~~~~\tilde\epsilon^{*}=-{2c_V\over\pi\tilde
Cl}.\end{equation}
By substitution of
the obtained values of the coupling constants into the action in
eq. (4.39), we come to the new conformal theory described by the following
perturbative action
\begin{equation}
S(\epsilon^{*},\tilde\epsilon^{*})=S_{WZNW}(g,l)~-~\epsilon^{*}
{}~\int d^2z~O(z,\bar
z)~-~\tilde\epsilon^{*}~\int d^2z~\tilde O(z,\bar z).\end{equation}
Apparently, each pair of the matrices $L,~\bar L$ solving the master equation
will lead to a CFT.

Let us consider the theory in eq. (4.51) with $\bar L=lI/(l-2c_V)$. The last is
the so-called Sugawara solution of the master equation. In
this case the perturbation in eq. (4.51) mimics well the perturbation term in
eq. (3.18). To make the resemblance between the two theories more transparent,
we present the nonunitary WZNW model in the form \cite{Soloviev-2}
\begin{equation}
S_{WZNW}(g,l)=S_{WZNW}(g,-l)~-~{2\over\pi l}~\int d^2z~:J^a\bar J^{\bar
a}\phi^{a\bar a}:,\end{equation}
where $S_{WZNW}(g,-l)$ is the action of the unitary conformal level $-l$ WZNW
model.

Then, the new perturbative conformal model can be rewritten as follows
\begin{equation}
S(\epsilon^{*},\tilde\epsilon^{*})=S_{WZNW}(g,-l)~-~{2\over\pi
(l-2c_V)}\left[\left(1-\frac{c_V}{\tilde
C}\right)\delta^{ab}~+~\left(\frac{c_V}{\tilde C}
-\frac{c_V}{C}\right)L^{(0)}_{ab}\right]~\int
d^2z~:J^a\bar J^{\bar b}\phi^{b\bar b}:.\end{equation}
Note that this is first order approximation to a certain exact conformal
field theory.

Now we can compare the perturbative conformal model with the theory given by
eq. (3.18). The two theories are renormalizable beyond conformal points. The
first one described by eq. (3.18) is renormalizable because it originates from
the renormalizable interaction of two WZNW models. The second theory is
renormalizable by construction. Apparently, in leading orders in $1/l$ they
coincide when $k=-l$ and
\begin{equation}
\tau\Sigma_{ab}={2c_V\over\pi
(l-2c_V)}\left[\left(\frac{1}{c_V}-\frac{1}{\tilde
C}\right)\delta^{ab}~+~\left(\frac{1}{\tilde C}
-\frac{1}{C}\right)L^{(0)}_{ab}\right].
\end{equation}
The last equation is, in fact, a condition of the conformal invariance of the
theory described by formula (3.18). In its turn, this condition will fix
conformal points of the system of two interacting WZNW models. The conformal
values of the Thirring coupling matrix $S_{a\bar a}$ are defined from the
following relation
\begin{equation}
S_{a\bar a}S_{b\bar a}={-c_V\over8k(k+2c_V)}\left[\left(\frac{1}{c_V}
-\frac{1}{\tilde
C}\right)\delta^{ab}~+~\left(\frac{1}{\tilde C}
-\frac{1}{C}\right)L^{(0)}_{ab}\right].
\end{equation}

In principle, it is possible to work out the
higher order corrections in $1/l$ to the
right hand side of eq. (4.55). But in this case, computations become much more
cumbersome. Already the lowest approximation displays that there exist as
many nontrivial conformal points in the theory of interacting WZNW models as
the number of
solutions to  the master Virasoro equation. Note that the conformal theory at
the conformal points given by eq. (4.55) is no longer based on the affine
symmetry, since the perturbation operators are affine descendants but not
affine primaries. Such a situation occures also in conformal representations of
the affine-Virasoro construction \cite{Obers-2}. However, it is still very
difficult
to identify the perturbative conformal model described by the action in eq.
(4.53) with one of many exact CFT's corresponding to the affine-Virasoro
construction. We only mention that the obtained perturbative conformal
model possesses the invariance under the so-called K-conjugation
\cite{Kiritsis} which amounts to the transformation given by eq. (4.38). The
classical action of the affine-Virasoro construction also enjoys this property
\cite{Yamron}.

All in all, we have proved that the theory of two interacting WZNW models with
the coupling matrix $S_{a\bar a}$ obeying the condition (4.55) is the infrared
limit of the perturbed nonunitary WZNW model described by eq. (4.51). In this
connection, it is worthwhile pointing out that among conformal solutions given
by eq. (4.55) there are diagonal ones which resemble closely the
conformal points of the generalized Thirring model found in \cite{Bardakci}.

While we have discussed the perturbation of the nonunitary WZNW model by two
orthogonal operators $O^{L,\bar L}$ and $O^{\tilde L,\bar L}$, there are two
more operators $O^{L,\bar{\tilde L}}$ and $O^{\tilde L,\bar{\tilde L}}$ which
are also appropriate for performing renormalizable perturbations. However, the
four-parameter perturbation around the nonunitary WZNW model, giving
new CFT's, do not correspond to the system of two interacting WZNW models.
Therefore, we do not consider perturbations of this type in the present paper.

\section{Conclusion}

Our main result is the equation (4.55) which gives the conditions of conformal
invariance for the theory of two interacting WZNW models. This equation
exhibits
that there are as many conformal points as solutions to the master-Virasoro
equation. Because it is very complicated, we could not identify the found
perturbative conformal points with known exact CFT's. One substantial fact is
that at the obtained conformal points the action of interacring WZNW models
possesses the invariance under
the K-conjugation. This permits us to believe that we deal actually with a
certain approximation to the quantum affine-Virasoro action. However, the
question still remains open.

\par \noindent
{\em Acknowledgement}: I would like to thank C. M. Hull for useful discussions.
I would also like to thank the PPARC and the European Commission Human Capital
and Mobility Programme for financial support.

\end{document}